\documentclass{article}
\usepackage{subcaption}
\usepackage{spconf,amsmath,graphicx}
\usepackage{hyperref,booktabs}

\title{Large Raw Emotional Dataset with Aggregation Mechanism}
\name{Vladimir Kondratenko$^{1}$ Artem Sokolov$^{1 3}$ Nikolay Karpov$^{1 2}$ Oleg Kutuzov$^{1}$ Nikita Savushkin$^{1}$ Fyodor Minkin$^{1}$}

\address{$^{1}$ Sber, Russia\\
$^{2}$ Nvidia, Armenia\\
$^{3}$ HSE University, Laboratory of Algorithms and Technologies for Network Analysis, Russia}
\begin{document}
%
\maketitle
\begin{abstract}
We present a new data set for speech emotion recognition (SER) tasks called Dusha. The corpus contains approximately 350 hours of data, more than 300 000 audio recordings with Russian speech and their transcripts. Therefore it is the biggest open bi-modal data collection for SER task nowadays. It is annotated using a crowd-sourcing platform and includes two subsets: acted and real-life. Acted subset has a more balanced class distribution than the unbalanced real-life part consisting of audio podcasts. So the first one is suitable for model pre-training, and the second is elaborated for fine-tuning purposes, model approbation, and validation. This paper describes pre-processing routine, annotation, and experiment with a baseline model to demonstrate some actual metrics which could be obtained with the Dusha data set.

\end{abstract}
\section{Introduction}
\label{sec:intro}


There are a lot of recent studies in the field of human behavior analysis and automatic speech emotion recognition (SER). Many of them use various inputs such as speech, video, and transcript as multi-modal data. The popular approach of such research is to invent a new neural network architecture and train it on the open data sets and benchmarks \cite{ser_deepml1}, \cite{ser_deepml2}. However, some aspects have a negative impact on the process of model training and evaluation. For instance, the small size of the open data aset frequently becomes a bottleneck for research. One more possible shortcoming is biasing between label annotation of data set and user emotions in the real world \cite{ser_trends}. It is highly desirable for a data set to involve as many label evaluators as possible but, practically, it is complicated enough to implement \cite{ser_annotation}. Another issue is the lack of speaker diversity which leads to the model underperforming when it faces a new speaker in a training set or in a real-time speech. 

These issues with the existing big open data sets motivated us to develop a new extensive database with Russian speech. We call it Dusha, which means Soul in Slavonic languages. It is designed to reveal such concepts as peace, openness and vast nature of the Eastern-European soul. We believe that our corpus can help to improve results in other languages using cross-corpus study \cite{milner2019cross} or transfer learning techniques on speech emotion recognition. The data set contains recordings of speech and their transcripts. That is why we call it bi-modal. 

Two sources of speech are used: acted crowd-sourced records and real-life podcasts in the Russian language. 
We consider that such a combination of domains is common in a real-life scenario when a model developer has less data from a target domain and much more from another crowd-sourced one.
We select the emotions that appear in the dialogue with a virtual assistant most frequently: Anger, Happiness, Neutral emotion, and Sadness.

Each item has been labelled by several annotators using 4 emotional classes so that markup could be aggregated into one confident label or multi-labelled. 
Along with the data, we share aggregation mechanism, so that any data scientist could get access to them to conduct research. 

This paper delivered to the open source an advanced speech emotion recognition data set with transcription. Also it describes approaches and methods for data set collection and markup. All data and processing scripts are released on a GitHub repository\footnote{\url{https://github.com/salute-developers/golos/tree/master/dusha}}.

\section{Related works}
\label{sec:format}

 To highlight our contribution, we analyzed existing Speech Emotional databases and compared our benchmarks with those including corpora with the Russian language. 

\subsection{Emotional Speech Datasets}
 
The interactive emotional dyadic motion capture database (IEMOCAP) \cite{iemocap} is a widely used multimodal data set that is de facto preferable for modern research comparison in emotion recognition and sentiment analysis. It contains visual data, audio tracks of dialogues, and transcribed text. Besides, this database includes motion data for faces and hands only. Five male and five female semi-professional actors recorded their voices for this data set. IEMOCAP exhibits the balanced distribution of emotions from the following list: happiness, anger, sadness, frustration, and neutral emotion. This material includes about 12 hours of an audio split in 5 dyadic sessions. Although the data set is balanced, its disadvantage is that it is not very extensive and has few speakers involved. Mostly, the benchmark is applicable for model comparing, yet it can cause an issue with precision during evaluation in live speech. 
It is a common researching practice to take a subset of IEMOCAP with four classes of emotions: happiness, sadness, anger and neutral emotion (where the excitement is combined with happiness) \cite{milner2019cross}. This set is referred to as IEMOCAP4.

The CMU Multimodal Opinion Sentiment and Emotion Intensity (CMU-MOSEI) database \cite{cmumosei} is another human multi-modal language benchmark. The data set is the next generation of CMU-MOSI \cite{mosi} and involves YouTube video recordings with the voices of 1000 distinct English speakers, text transcription of audio, and emotion annotation for each utterance. In addition to the size of CMU-MOSEI, one of its strong points is that emotions are not acted. However, the emotion annotation of this benchmark was conducted by only 3 crowdsourced persons. Potentially, such a few number annotators could lead to a gap in accuracy for the evaluation and include some bias compared to real data, even if they pass special training.

Among widely-spoken languages, Chinese (Mandarin) and Spanish are also covered by numerous data sets. German domain is widely represented in emotion databases too \cite{german2}, \cite{german3}. The most famous one is EmoDB \cite{emodb}.

An attempt to create an enormous repository by joining several various languages was described in \cite{emonet}. The authors presented a united database that included subsets with English, German, Chinese, Turkish and other languages.

\subsection{Datasets in Russian Domain}
Currently, there are very few data collections for emotional speech recognition available in the Russian language.

One of the first attempts to organise a Russian emotional data set is described in \cite{ruslana}. This set of audio utterances and their transcriptions is called Russian Language Affective speech database (RUSLANA). Students of various Russian universities, participating as speakers, dictated in total 6.400 utterances with the corresponding emotions.

Russian Multimodal Corpus of Dyadic Interaction for Studying Emotion Recognition (RAMAS) \cite{ramas} is another widely known Russian language data set. Similar to IEMOCAP, it includes acted recordings with 7 hours of emotional speech. The corpus provides video and audio modality, transcripts, motion, and physiology data. It annotated the following emotions: Anger, Sadness, Disgust, Happiness, Fear, Surprise. Ten actors participated in the recording of the video clips for this benchmark.

One more Russian database which could be employed for SER is Multimodal Russian Corpus (MURCO) \cite{murco} which is a part of the Russian National Corpus (RNC). It stores clips from Russian cinematography, TV and radio programs, recordings of usual conversations, etc. Although MURCO has millions of recordings, it has quite obsolete and unfriendly interfaces for automatic data retrieving. The complete list of emotion classes is not defined. 

We consider the problem of large-scale data sets for SER tasks.
When faced with real-life emotions, the data set would become a framework to conduct research and establish a connection between obtained results in the laboratory and system behavior.
In addition, MLS \cite{pratap2020mls} and Golos \cite{karpov2021golos} data sets play a major part in the automatic speech recognition (ASR) task.
Therefore, we decided to collect and share a large multimodal (audio and text) data set in the Russian domain and involve both acted and real-life data.

\section{Data Acquisition}

The Dusha data set consists of two logical parts which are obtained in completely different ways. The first one is collected with the assistance of non-professional actors on a popular crowd-sourcing platform Yandex Tolloka\footnote{\url{https://toloka.yandex.ru}}. 
Further in the text, we call it “\textit{Crowd domain}” or "\textit{Crowd}". 
The second part consists of a speech from various emotional podcasts. We call it "\textit{Podcast domain}" or "\textit{Podcast}". 

\subsection{Crowd subset collection}
The text for crowd recordings was chosen from genuine requests which users fulfilled via virtual voice assistant Salute and SmartSpeech \footnote{\url{https://github.com/salute-developers/salute-speech}} service for speech recognition. 
Raw data set included tens of millions of recordings and their transcriptions. It is evident that most voice requests involve an urge to do something like "Salute, turn on YouTube", "Salute, sign me up for a hairdresser" and other phrases and talks which users send to their voice assistant with neutral emotion. To balance our data, we filtered out requests and kept recordings with conversation (chit chat) because this subset could include more explicit emotional utterances. To do so we employed Salute 
internal intent classifier, which separates various types of voice commands and selects chatter requests when no action except response is required. The resulting subset was several millions of utterances.

Next, we applied an emotional pseudo labelling of texts to establish what emotions could be acted for utterances. 
We employed a simple classifier on the top of a BERT-large version of well-known BERT architecture \cite{devlin2018bert} which was trained from scratch internally and could classify our texts for 4 target sentiments: anger, happiness, sadness and neutral emotion.
The investigation result demonstrates that neutral emotion dominated in a significant number of cases.
To evaluate our pseudo labels we conducted a survey on a crowd-sourcing platform where we asked to label manually a small part (\~{}10.000) of utterances and compare with classifier results. It shows that our pseudo labels are sufficiently accurate.
We use them to sample emotional utterances and decrease the count of neutral recordings.

Next, we carried out audio voicing with the help of non-professional actors on a crowd-source platform. We took pseudo labels predicted on the previous step into account and for each phrase we set one emotion from the label and one more with similar emotion valence or neutral sentiment. For instance, we organized emotions in pairs positive/neutral, sadness/neutral, anger/sadness etc.

Thus, the actors had to pronounce the text with one of the emotions from the pair. Also, we provided a description on how to better voice the emotion.

Totally, we obtained 201 850 acted emotions with 2 068 unique speakers where, neutral emotion dominates as in real-life situations however other classes are quite balanced. Blue column on Figure~\ref{fig:len_and_emotion_distrib} \textbf{(a)} represents the time length distribution. As people used their own equipment, the quality of audio files differs. Audio can contain background noises, such as children and animal voices or street sounds. Total length is about 255 hours.

\subsection{Podcast subset collection}
The Podcast subset was designed to diversify data in the Dusha database. Emotions in these recordings are not performed, but rather sincere. Furthermore, the distribution of emotions for this data set corresponds better to their distribution in usual human speech. \textit{Podcast domain} is not balanced and the neutral emotion class substantially outnumbers other classes. Moreover, since acted emotions may differ slightly from the spontaneous real-life emotions, we consider it reasonable to keep this subset with natural class distribution in the Dusha. The Podcast could be used for fine-tuning goals and assessing the quality of the model for the production system. 

\begin{figure}[t]
  \centering
  \begin{subfigure}[b]{0.45\linewidth}
    \centering
    \includegraphics[width=\linewidth]{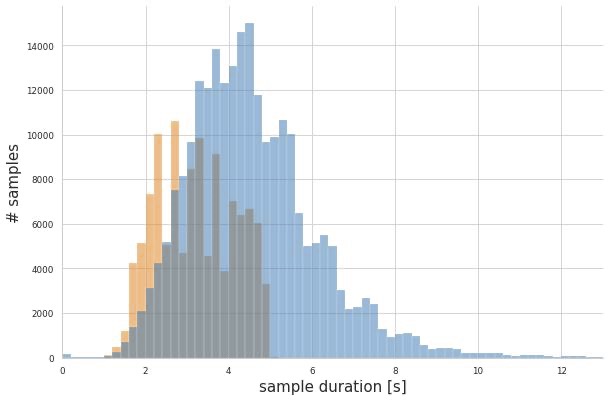}
    \caption{Audio length distribution on Dusha corpus domain. Blue - Crowd domain, Orange - Podcast domain.}
  \end{subfigure}
  \hfill
  \begin{subfigure}[b]{0.45\linewidth}
    \centering
    \includegraphics[width=\linewidth]{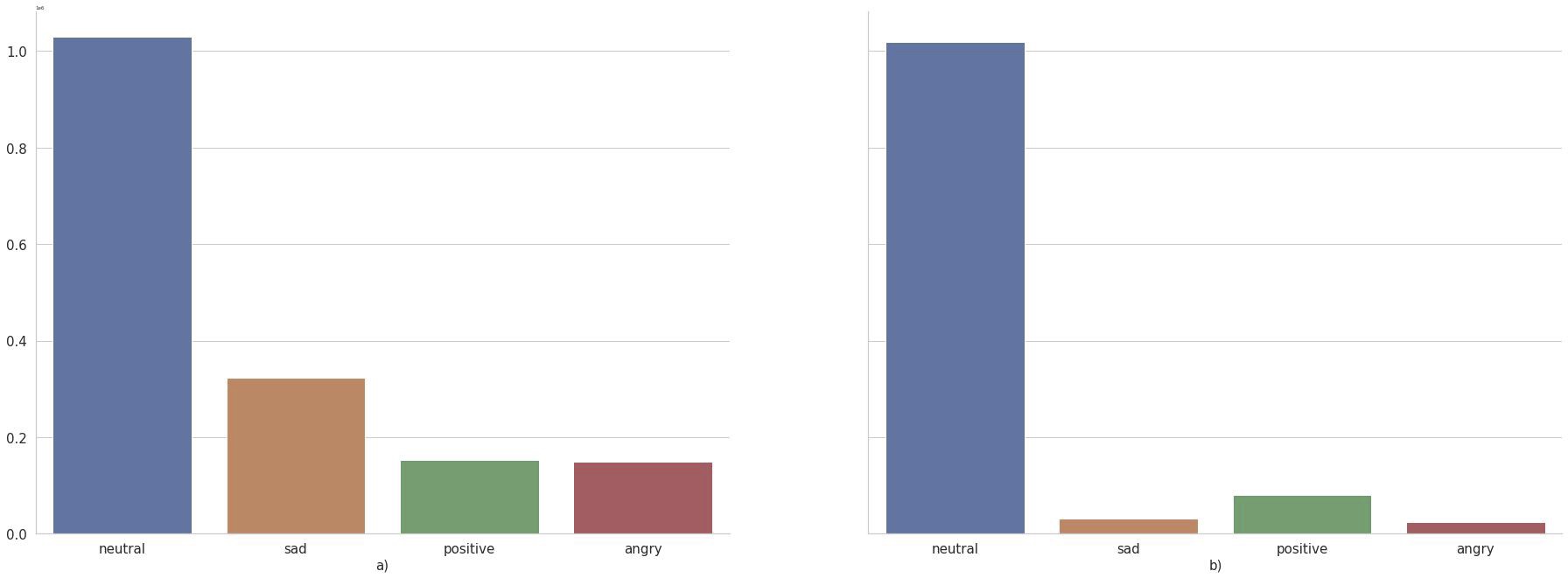}
    \caption{Emotion distribution in Dusha corpus. Blue - Neutral, Orange - Sad, Green - Positive, Red - Angry.}
  \end{subfigure}
   \caption{}
  \label{fig:len_and_emotion_distrib}
\end{figure}

We obtained a topic diversity and included entries on politics, IT, games, relationships, etc. We do not fulfil any specific podcast choosing or filtering and just trying to cover various conversation topics. Recordings were sliced into 5-second segments by a voice activity detector (VAD) to simplify emotion annotation (See Figure~\ref{fig:len_and_emotion_distrib}(\textbf{a}) orange color). 
A total of 6240 podcasts were used, of which 102 113 samples were selected. In general, the Podcast audio is recorded with professional equipment and has a better quality than the Crowd. We normalized files to 16-bit, 16 000 Hz. 
Total length is greater than 90 hours.

\subsection{Post-processing and annotation}
To avoid implicit bias in annotation on crowd-sourcing platform each person took the training and passed the exam. All participants who had attained a passing score above 80\% were allowed to evaluate data.

Participants listened to the audio only and did not have access to the transcript to evaluate emotions of \textit{Crowd} and \textit{Podcast domain}. Annotators were given instructions to choose their labels using one of the five options:

\textbf{Positive:} the text is spoken with a smile, or laughter, or admiration, or a playful tone, or there are pronounced stresses on words emphasizing the positive.

\textbf{Neutral:} the voice is still and calm, there is no emotion in the voice. At the same time, even if the text is clearly negative (for example, “how tired you are”) or positive (for example, “what a fine fellow you are”), this emotion is not expressed in voices, it is necessary to mark the emotion as neutral.

\textbf{Sadness:} the text is pronounced with sadness, melancholy, a faded voice.

\textbf{Anger/Irritation:} if the text is spoken with anger or annoyance, or the user is yelling or speaking through gritted teeth, or there are pronounced stresses on words emphasizing the negative.

\textbf{Other:} the recording is too quiet, hissing, rattling or there is no voice.

In order to ensure the quality of markup, each person from time to time got a control task in which we knew the correct label. We named such control tasks "honeypots". If an answer to the control task was correct he or she would continue to mark up. During annotation 303 963 recordings were evaluated and 1 715 301 emotion labels were accumulated. 

\section{Data set Overview}

\subsection{Raw Data set}
Our raw metadata includes at least three labels given by independent annotators per sample and several fields for pure emotional markup without any aggregation. 
Independent annotators have an independent opinion about emotion labels. In case of disagreement more people were involved to mark one sample. 


A list of fields of raw metadata is provided below:
\textbf{wav\_path} - relative path to audio file;
\textbf{annotator\_id} - unique id of annotator;
\textbf{annotator\_emo} - emotion mark given by annotator;
\textbf{annotator\_emo} - emotion mark given by annotator;
\textbf{golden\_emo} - emotion mark of control tasks (honeypots);
\textbf{speaker\_text} - original speaker text to pronounce;
\textbf{speaker\_emo} - intentional emotion of the audio;
\textbf{source\_id} - unique id of speaker or podcast;

Metadata stores information about all applicable emotions to each recording, voting results and other specific data. It enables researchers to explore consistency of markup and try various methods to customise markup for data sampling with specific annotation confidence level.
In order to get data set for machine learning purposes we have to group labels by audio files and aggregate into single-labels or multi-labels. We call this "aggregation" mechanism.
For aggregation of raw data we use Dawid-Skene (DS) algorithm \cite{dawid1979maximum} with confidence threshold to limit the level of agreement. 
We choose an empirically selected threshold 0.9. Unlike raw corpus, subset we get could be employed for SER model.

The emotion distribution per domain of aggregated annotation are depicted on Figure~\ref{fig:len_and_emotion_distrib}(\textbf{b}) and Table~\ref{tab:annotation_results}. A list of fields of this metadata is provided below:
\textbf{wav\_path} - relative path to audio file;
\textbf{emotion} - aggregated emotion mark;
\textbf{speaker\_text} - original text in the audio record;
\textbf{speaker\_emo} - intentional emotion of the audio;
\textbf{source\_id} - unique id of speaker.
The number of items and duration in the aggregated training and test subsets are represented in Table~\ref{tab:dataset_separation}. 

\begin{table}[t!]
  \caption{Emotion Files Distribution After Aggregation Mechanism using Dawid-Skene algorithm with threshold 0.9.}
  \label{tab:annotation_results}
  \centering
  \resizebox{\columnwidth}{!}{%
  \begin{tabular}{lrrrrrr}
    \toprule
    \textbf{Domain} & \textbf{Pos} & \textbf{Sad} & \textbf{Ang} & \textbf{Neu} & \textbf{Oth} & \textbf{Total} \\
    \midrule
    Crowd & 15446 & 23316 & 17120 & 106850 & 1655 & 164387 \\
    Podcast & 5909 & 1170 & 2057 & 81104 & 222 & 90462 \\
  \end{tabular}
  }
\end{table}

\begin{table}[t!]
  \caption{Amount of Data After Aggregation Mechanism using Dawid-Skene algorithm with threshold 0.9.}
  \label{tab:dataset_separation}
  \centering
  \resizebox{\columnwidth}{!}{%
  \begin{tabular}{ l|r r|r r }
    \toprule
    \multicolumn{1}{l|}{\textbf{Domain}} & \multicolumn{2}{c|}{\textbf{Training files and hours}} & \multicolumn{2}{c}{\textbf{Test files and hours}} \\
    \midrule
    Crowd  & 150352 & 188 h. 44 min.  & 14035 & 18 h. 29 min. \\
    Podcast & 79825 & 71 h. 23 min. &  10637 & 09 h. 25 min. \\
    \bottomrule
    Total  & 230177 & 260 h. 07 min. & 24672 & 27 h. 54 min. \\
  \end{tabular}
  }
\end{table}

\begin{table}[t!]
  \caption{Experiment Results on Dusha Benchmark}
  \label{tab:exps_result}
  \centering
  \resizebox{\columnwidth}{!}{%
  \begin{tabular}{l|ccc|ccc}
  \toprule
  \multicolumn{1}{l|}{ } & \multicolumn{3}{c|}{\textbf{Crowd test}} & \multicolumn{3}{c}{\textbf{Podcast test}} \\
  \textbf{Training setup} & \textit{UA} & \textit{WA} & \textit{F1} & \textit{UA} & \textit{WA} & \textit{F1} \\
  \midrule
  Dusha & \textbf{0.83} & 0.76 & 0.77 & \textbf{0.89} & 0.53 & 0.54 0.01 \\
  \toprule
  \end{tabular}
  }
\end{table}

\subsection{Baseline Implementation Details}
We conduct experiments using the shallow baseline model in order to simplify the entry threshold for researchers who will benchmark using our data set.

We use common metrics for SER tasks: macro F1 score (\textit{F1}), Unweighted Accuracy (\textit{UA}), Weighted Accuracy (\textit{WA}). These validation metrics are calculated on Crowd and Podcast testing sets, which are created using Dawid-Skene algorithm with confidence $>$ 0.9. 

We train a baseline model from scratch with both Dusha parts (\textbf{Crowd} and \textbf{Podcast}). 
Additionally, we train our baseline model on IEMOCAP4 to compare it with other state-of-the-art (SOTA) solutions for speech emotion recognition.

For our experiments we employ an audio modality only. 
As input we pass 64 Mel-filterbank calculated from 20ms windows with a 10ms overlap. 
Next, features are received at a simple MobileNetV2\cite{mobilenetv2} based architecture with a self-attention layer described in SAGAN\cite{zhang2019self}. 
Input Mel features are passed through a sequence of inverted residual blocks as it is done in \cite{mobilenetv2}, but with custom layers configuration.
Then we apply a convolutional self-attention layer followed by a global average pooling.
After that, we pass the resulting vector (one number for each feature map) through a fully connected layer to get classification results.

The model is implemented in Pytorch, using the Adam\cite{adam} optimizer with learning rate $0.001$, a weight decay of $10^{-6}$ and without gradient clipping. We train models 100 epochs with batch size 64.

\subsection{Benchmark Results}
The results of our experiments are presented in Table~\ref{tab:exps_result}. For all test subsets \textit{UA} is higher than \textit{WA}. It could be explained by the neutral emotion dominance. The corpus includes emotion distribution as people faced it. However each researcher or engineer can filter out emotions as he/she wants.

Our baseline model trained on IEMOCAP4 subset of IEMOCAP shows $0.59 \pm 0.01$ unweighted accuracy \textit{UA}, $0.59 \pm 0.01$ weighted accuracy \textit{WA}, and $0.59 \pm 0.01$ \textit{macro F1} score with 5 sessions cross testing. Actual SOTA result we showed with IEMOCAP were considerably better, but we didn't set the goal to obtain the best metrics. We demonstrated abilities of the utilized architecture for the popular data set.

\section{Conclusion}
In this study, we introduce in details the novel speech data set for emotion recognition called "Dusha". The data has been taken from two different sources. The first one is 255 hours of audio with text transcriptions. This is an acted subset obtained and labeled via a crowd-sourcing platform. The second subset is taken from various podcasts and its size is about 90 hours.

The distinctive feature of Dusha is that we provide a raw emotional data set and an example of an aggregation mechanism. The Dusha's markup can be aggregated into single-labels or multi-labels. The research community can use our example of a label aggregation or set-up in their own experiments with customized filtering.
We open-sourced a code to benchmark models using Dusha and conduct an experiment with baseline model to demonstrate obtained metrics with default emotion distribution.

\section{Acknowledgments}
The work of Artem Sokolov is partially supported by RSF (Russian Science Foundation) grant 20-71-10010.

\bibliographystyle{IEEEbib}
\bibliography{refs}

\begin{thebibliography}{10}

\bibitem{ser_deepml1}
Panagiotis Tzirakis, Zhang Jiehao, and Bjorn~W. Schuller.,
\newblock ``End-to-end speech emotion recognition using deep neural networks,''
\newblock {\em IEEE international conference on acoustics, speech and signal
  processing (ICASSP)}, 2018.

\bibitem{ser_deepml2}
Wei-Cheng Lin and Busso. Carlos,
\newblock ``An efficient temporal modeling approach for speech emotion
  recognition by mapping varied duration sentences into fixed number of
  chunks,''
\newblock {\em Intespeech}, 2020.

\bibitem{ser_trends}
Björn~W. Schuller,
\newblock ``Speech emotion recognition: Two decades in a nutshell, benchmarks,
  and ongoing trends,''
\newblock {\em Communications of the ACM}, 2018.

\bibitem{ser_annotation}
Laurence Devillers, Vidrascu Laurence, and Lamel. Lori,
\newblock ``Challenges in real-life emotion annotation and machine learning
  based detection,''
\newblock {\em Neural Networks}, 2005.

\bibitem{milner2019cross}
Rosanna Milner, Md~Asif Jalal, Raymond~WM Ng, and Thomas Hain,
\newblock ``A cross-corpus study on speech emotion recognition,''
\newblock in {\em 2019 IEEE Automatic Speech Recognition and Understanding
  Workshop (ASRU)}. IEEE, 2019, pp. 304--311.

\bibitem{iemocap}
Busso Carlos, Bulut Murtaza, Lee Chi-Chun, Kazemzadeh Abe, Mower Emily, Kim
  Samuel, N.~Chang Jeannette, Lee Sungbok, and S~Narayanan. Shrikanth,
\newblock ``Iemocap: Interactive emotional dyadic motion capture database,''
\newblock {\em Language resources and evaluation}, 2008.

\bibitem{cmumosei}
Zadeh Amir, Liang Paul, Pu, Vanbriesen Jonathan, Poria Soujanya, Tong Edmund,
  Cambria Erik, Chen Minghai, and Louis-Philippe Morency.,
\newblock ``Multimodal language analysis in the wild: Cmu-mosei dataset and
  interpretable dynamic fusion graph,''
\newblock {\em Proceedings of the 56th Annual Meeting of the Association for
  Computational Linguistics (Long Papers)}, 2018.

\bibitem{mosi}
Amir Zadeh, Rowan Zellers, Eli Pincus, and Louis-Philippe Morency.,
\newblock ``Mosi: multimodal corpus of sentiment intensity and subjectivity
  analysis in online opinion videos,'' 2016.

\bibitem{german2}
Florian Schiel, Silke Steininger, and Ulrich T{\"u}rk,
\newblock ``The smartkom multimodal corpus at bas.,''
\newblock in {\em LREC}. Citeseer, 2002.

\bibitem{german3}
Bjorn Schuller, Arsic Dejan, Rigoll Gerhard, Wimmer Matthias, and Radig. Bernd,
\newblock ``Audiovisual behavior modeling by combined feature spaces,”in
  proceedings of the,''
\newblock {\em International Conference on Acoustics, Speech and Signal
  Processing (ICASSP))}, 2007.

\bibitem{emodb}
Felix Burkhardt, Paeschke Astrid, Rolfes Miriam, F.~Sendlmeier Walter, and
  Weiss. Benjamin,
\newblock ``A database of german emotional speech,''
\newblock {\em Interspeech)}, 2005.

\bibitem{emonet}
Maurice Gerczuk, Amiriparian Shahin, Ottl Sandra, and Bjorn~W. Schuller.,
\newblock ``Emonet: A transfer learning framework for multi-corpus speech
  emotion recognition,''
\newblock {\em IEEE Transactions on Affective Computing)}, 2021.

\bibitem{ruslana}
Veronika Makarova and Valery~A. Petrushin,
\newblock ``Ruslana: A database of russian emotional utterances,''
\newblock {\em Seventh international conference on spoken language processing},
  2002.

\bibitem{ramas}
Olga Perepelkina, Evdokia Kazimirova, and Maria. Konstantinova,
\newblock ``Ramas: Russian multimodal corpus of dyadic interaction for
  affective computing,''
\newblock {\em International Conference on Speech and Computer}, 2018.

\bibitem{murco}
Svetlana Savchuk and Alexandra. Makhova,
\newblock ``Multimodal russian corpus and its use in emotional studies,''
\newblock {\em Russian Journal of Communication}, 2021.

\bibitem{pratap2020mls}
Vineel Pratap, Qiantong Xu, Anuroop Sriram, Gabriel Synnaeve, and Ronan
  Collobert,
\newblock ``Mls: A large-scale multilingual dataset for speech research,''
\newblock {\em arXiv preprint arXiv:2012.03411}, 2020.

\bibitem{karpov2021golos}
Nikolay Karpov, Alexander Denisenko, and Fedor Minkin,
\newblock ``Golos: Russian dataset for speech research,'' 2021.

\bibitem{devlin2018bert}
Jacob Devlin, Ming-Wei Chang, Kenton Lee, and Kristina Toutanova,
\newblock ``Bert: Pre-training of deep bidirectional transformers for language
  understanding,'' 2018.

\bibitem{dawid1979maximum}
Alexander~Philip Dawid and Allan~M Skene,
\newblock ``Maximum likelihood estimation of observer error-rates using the em
  algorithm,''
\newblock {\em Journal of the Royal Statistical Society: Series C (Applied
  Statistics)}, vol. 28, no. 1, pp. 20--28, 1979.

\bibitem{mobilenetv2}
Mark Sandler, Andrew~G. Howard, Menglong Zhu, Andrey Zhmoginov, and
  Liang{-}Chieh Chen,
\newblock ``Inverted residuals and linear bottlenecks: Mobile networks for
  classification, detection and segmentation,'' 2018.

\bibitem{zhang2019self}
Han Zhang, Ian Goodfellow, Dimitris Metaxas, and Augustus Odena,
\newblock ``Self-attention generative adversarial networks,''
\newblock in {\em International conference on machine learning}. PMLR, 2019,
  pp. 7354--7363.

\bibitem{adam}
Diederik~P. Kingma and Jimmy Ba,
\newblock ``Adam: {A} method for stochastic optimization,''
\newblock in {\em 3rd International Conference on Learning Representations,
  {ICLR} 2015, San Diego, CA, USA, May 7-9, 2015, Conference Track
  Proceedings}, Yoshua Bengio and Yann LeCun, Eds., 2015.

\end{thebibliography}

\end{document}